# Study of the mechanism of electroacupuncture regulating ferroptosis, inhibiting bladder neck fibrosis, and improving bladder urination function after suprasacral spinal cord injury using proteomics

Jin-Can Liu[a†], Li-Ya Tang[a†], Xiao-Ying Sun[a], Qi-Rui Qu[a], Qiong Liu[a], Lu Zhou[b], Hong Zhang[a], Bruce Song[c], Ming Xu[a], Kun Ai[a*].

a College of Acupuncture, Massage and Rehabilitation, Hunan University of Chinese Medicine, Changsha, Hunan 410208, China.
b Department of rehabilitation medicine, Chenzhou first people's Hospital, Chenzhou, Hunan 423000, China.
c Key Laboratory of Sports and Physical Health Ministry of Education, Beijing Sport University Bejing 100084, China
[†]The author contributes to this paper.

[*]**Corresponding authors:**

Professor Kun Ai, Hunan University of Chinese Medicine, No. 300, Xueshi Road, Changsha, Hunan, China, 410208. Email: aikun650@qq.com.

**Abstract**

**Purpose** The aim of this study was to explore whether electroacupuncture regulates phenotypic transformation of smooth muscle cells by inhibiting ferroptosis and inhibiting fibrosis, thereby improving bladder urination function after suprasacral spinal cord injury (SSCI).

**Methods** The experiment was divided into sham, model, and electroacupuncture group. After 10 days of electroacupuncture intervention, urodynamic examination was performed, and bladder neck was taken for HE staining, tandem mass tag (TMT)-based quantitative proteomics analysis, Western blot(WB) detection, ferrous ion concentration detection and Masson staining.

**Results** Urodynamic results showed that electroacupuncture could improve the micturition function of NB rat bladders, and HE staining showed that pathological changes, such as inflammation of bladder neck tissue, were improved after electroacupuncture intervention. Proteomics and biological information analysis show that ferroptosis and the contractile smooth muscle cell marker SMTN were significantly enriched ($P<0.05$). WB analysis showed that after electroacupuncture treatment, the expressions of STEAP3, TF, TFRC, DMT1 and P53 decreased significantly ($P<0.05$), while the expression of SMTN increased significantly ($P<0.05$). The ferrous ion concentration detection results showed that the ferrous ion concentration in the bladder neck tissue decreased significantly after electroacupuncture treatment ($P<0.01$). Masson's trichrome staining showed that the deposition of collagen fibers in the bladder neck muscle layer decreased significantly after electroacupuncture treatment ($P<0.01$).

**Conclusion** Electroacupuncture may prevent the phenotype of bladder neck smooth muscle cells from transforming from contraction type to synthesis type by inhibiting ferroptosis, inhibit bladder neck fibrosis, improve bladder neck compliance, and thus improve bladder urination function after SSCI.

# 1. Introduction

Urination is controlled by the detrusor muscle, bladder neck (internal urethral sphincter), and external urethral sphincter. Under normal circumstances, when the bladder is full, under the control of the higher urination center, the parasympathetic nerve (S2-S4) innervating the detrusor muscle of the bladder is activated, causing the detrusor muscle to contract; the sympathetic nerve innervating the bladder neck (T11-L2) and the somatic nerve innervating the external urethral sphincter (Onuf's nucleus of S2-S4) are inhibited; and the bladder neck and external urethral sphincter dilate, causing urination [1].

After suprasacral spinal cord injury (SSCI), sympathetic and parasympathetic coordination is disrupted due to the loss of upper center control of the lower center of the spinal cord, resulting in the failure of the bladder neck or external urethral sphincter to relax during urination, leading to the detrusor-sphincter coordination disorder neurogenic bladder (NB). Therefore, whether the bladder neck and external urethral sphincter can synergistically dilate during urination is currently the focus of clinical research. Most studies mainly focus on the external urethral sphincter, and relatively few studies have focused on the bladder neck.

After spinal cord injury, the sympathetic innervation area of the bladder neck is overexcited, and the bladder neck continues to contract, resulting in bladder tissue ischemia and hypoxia. This causes changes in gene and protein levels, disrupting metabolism in smooth muscle cells and aggregating extracellular matrix components such as collagen and glycoprotein, leading to tissue fibrosis [2–5]. Smooth muscle cell fibrosis reduces bladder compliance. In addition, an increase in urethral orifice pressure during urination is an important factor that aggravates the disturbance of urination caused by detrusor sphincter dysfunction. Therefore, the reduction in compliance caused by smooth muscle fibrosis of the bladder neck is an important factor affecting NB voiding disorders after spinal cord injury. Further research on preventing smooth muscle cell fibrosis in the bladder neck can provide a new approach for the clinical treatment of urinary voiding disorders after SSCI.

It has been reported that relaxin [6, 7] and botulinum toxin A [8] are often used clinically to inhibit smooth muscle and striated muscle contraction, and these interventions can also effectively inhibit smooth muscle fibrosis of the bladder [9, 10]. Clinical and experimental studies have shown that acupuncture is effective in the treatment of NB after SSCI, can inhibit cystic fibrosis, and improve bladder compliance; it has been widely used in clinical practice because of its advantages of simplicity, lack of side effects, and low cost [11, 12]. This study provides theoretical support for the clinical treatment of NB after SSCI by exploring potential therapeutic targets.

Tandem mass tag (TMT)-based quantitative proteomics technology can reveal the differentially expressed proteins (DEPs) in the life process or disease occurrence, help to discover the potential physiological and pathological mechanisms of the body, and is an effective method to suggest potential therapeutic targets of drugs [13, 14]. Due to its unique advantages, it has been widely used in TCM research to explore complex regulatory mechanisms [15–17].

Studies have found that the phenotypic transformation of smooth muscle cells in the bladder is one of the causes of smooth muscle fibrosis [18, 19], and ferroptosis can regulate the phenotypic transformation of smooth muscle cells [20, 21]. This study used TMT quantitative proteomics technology to investigate the possible mechanism of electroacupuncture in the treatment of NB after SSCI. The results of bioinformatics analysis showed that ferroptosis and its key molecular, STEAP3 are significantly enriched.

Ferroptosis is a form of regulatory cell death characterized by iron-dependent oxidative damage, plasma membrane rupture, and the release of injury-related molecular patterns [22, 23]. Relevant studies have found that ferroptosis can mediate the transformation of smooth muscle cells from contractile to synthetic, secrete a large amount of extracellular matrix, and eventually lead to fibrosis [24, 25]. To further explore this potential relationship, we verified the relevant key proteins in the ferroptosis signaling pathway using Western blotting (WB), detected the concentration of ferrous ions using a biochemical kit, and observed pathological changes in the bladder neck with Masson's trichome staining.

This study aimed to explore whether electroacupuncture can regulate the phenotypic transformation of smooth muscle cells by mediating ferroptosis of the smooth muscle of the bladder neck, inhibiting fibrosis, improving bladder compliance, improving urination dysfunction, and providing a scientific basis for electroacupuncture treatment of NB after SSCI.

## 2. Materials and methods
### 2.1 Animal experimental design
#### 2.1.1 Source and grouping of experimental animals

Fifty-two healthy adult specific pathogen-free (SPF) Sprague-Dawley (SD) rats (250–280 g) were selected and provided by the Animal Laboratory Center of Hunan University of Traditional Chinese Medicine (certificate no.: 1107271911006889). Rats were housed in cages under standard conditions (temperature 24–26 ºC, humidity 50–70%) at the Animal Center Laboratory of Hunan University of Chinese Medicine. The rats were randomly divided into sham operation (12 rats) and modeling groups (40 rats). Hassan Shaker spinal cord transection method was used to make SSCI models of the rats in the modeling group [26]; 34 rats survived after spinal shock, of which 24 met the modeling standards. They were randomly divided into Model (n = 12) and Model+electroacupuncture (EA) groups (n = 12).

#### 2.1.2 Modeling

After seven days of routine feeding, the rats were deprived of water for 24 h. Two hours before modeling, 200,000 units of penicillin sodium (North China Pharmaceutical, Shijiazhuang, Hebei, China) were injected intraperitoneally for anti-inflammatory purposes. Pentobarbital sodium (3 %; Merck KGaA, San Jose, CA, USA) was administered intraperitoneally. After thorough anesthesia, the rats were fixed on the mouse plate in a prone position, with the floating ribs connecting the 13th thoracic vertebra as a bony marker and positioned upward at the spinous process of the 8-9 thoracic vertebrae. A 2–3 cm-long longitudinal incision was made to expose the T8 and T9 spinous processes and the adjacent vertebral arch. A micro rongeur was used to remove the T8 lamina from the tail to the lateral pedicle and expose the spinal cord. The spinal cord was then pulled out of the intervertebral space using a dental hook, and the tip was quickly cut with a surgical knife. Finally, forceps were used to gently lift both ends of the spine to ensure complete transection of the spinal cord. Finally, the muscle was sutured, the incision and surrounding area were disinfected with 5% iodophor, and the skin was sutured. In the sham operation group, only the skin was cut, and the muscles were separated and sutured.

After the operation, the rats were placed on a constant-temperature electric blanket until they woke up and were fed in a single cage. Penicillin sodium was injected intraperitoneally one week after surgery. The Crede technique was used to assist with urination every morning, afternoon, and evening. Bedsores and self-injuries were disinfected using iodophors.

**2.1.3 Animal model evaluation**

Evaluation of hind limb motor function: The hind limbs of the rats were dragged when walking, and the Basso-Beattie-Bresnahan (BBB) motor score [27] was 0, which is considered a successful model of SSCI. Assessment of bladder urination function: After spinal cord shock, rats with NB-type bladders showed bladder distension and mild moisture in the lower abdomen and cage. Resistance to the urethral opening was observed during manual urination. Urine flow mechanics showed that the leak point pressure (LPP) and maximum cystometric capacity (MCC) of the bladder increased, indicating successful modeling.

**2.1.4 Electroacupuncture intervention**

Twelve rats in the Model+EA group were given electroacupuncture treatment after the stage of spinal shock (19th day after modeling) [28] using the SDZ-V type Huatuo brand electroacupuncture therapy instrument (Suzhou Medical Supplies Factory Co., LTD.): The Ciliao (BL32) (opened 5–10 mm at the upper median of spinous process space of the second and third sacrum), Zhongji (RN3) (the intersection of upper 3/4 and lower 1/4 of the line between sternoclavicular union and the upper margin of the symphysis pubis was positioned as Shenque point, and the intersection of upper 4/5 and lower 1/5 of the line between Shenque point and the upper margin of the symphysis pubis was positioned as the Zhongji), and Sanyinjiao (SP6) (1 cm directly above the tip of the inner ankle of the hind limb). The rats were fixed on a mouse rack and treated with 30 1-inch needles for 15 mm of Ciliao (BL32), 5 mm of Zhongji (RN3), and 5 mm of Sanyinjiao (SP6) once a day for 20 min each for 10 days. The model and sham operation groups were restrained in the same manner for 20 min, but no electroacupuncture intervention was performed on the sham operation group.

**2.2 Urodynamics tests**

After 10 days of electroacupuncture intervention, the urine flow dynamics of the rats in each group were detected. The rats were injected intraperitoneally with 3% pentobarbital sodium and placed in the supine position after thorough anesthesia. A longitudinal incision was made on the anterior median line of the symphysis pubis to expose the bladder, a small incision was made at the apex of the bladder, and an F3 catheter was inserted into the bladder. The multi-channel physiological recorder (Biopac, USA), micro-infusion pump (Zhejiang Smiths Medical Instrument Co., LTD.), and F3 catheter were connected through a three-way pipe, and the micro-infusion pump was opened to inject normal saline at a temperature ranging from 25 °C–35 °C and at a speed of 0.1 ml/min. LPP and MCC were recorded.

**2.3 Hematoxylin and eosin staining**

The left ventricle was injected with 0.9% normal saline until clear liquid flowed from the right auricle. The bladder tissue was bluntly separated, and the bladder neck was removed on ice, fixed with 0.4% paraformaldehyde, embedded in paraffin, sliced, dewaxed with xylene, washed with ethanol, stained with hematoxylin (Servicebio, China) for 10 min, and soaked in tap water for 15 min. The sections were stained with 1% eosin (Servicebio) for 10 min, dehydrated with ethanol and xylene, and fixed with neutral gum tablets. The bladder neck structure was observed under a microscope.

**2.4 Tandem mass tag (TMT) quantitative proteomics analysis**

### 2.4.1 Preparation of protein samples

Three rat bladder necks were collected from each group for proteomic analysis, and bladder neck tissues were extracted under deep anesthesia with 3% pentobarbital sodium. Each sample was added to 1000 μL working liquid (RIPA lysate was mixed with a protease inhibitor and pre-cooled on ice to obtain the working liquid) and then fully dissolved by ultrasound in an ice bath (5 min). It was centrifuged at 4 °C for 15 min, 14000g, and the supernatant was transferred to a new Eppendorf (EP) tube. BCA protein assay kit (Elabscience Biotechnology Co., Ltd; E-BC-K318-M) was used to quantitatively determine protein concentration according to instructions.

### 2.4.2 Tandem mass tag (TMT) quantitative labeling

For each sample, 100 μg supernatant was reduced, alkylated, acetone precipitated, and protein redissolved to obtain the peptides of the corresponding sample. Different TMT solutions (20 μL) were absorbed into the corresponding sample, mixed, centrifuged, and incubated at room temperature for 1 h. Hydroxylamine was added to 100 mm of the sample, and the reaction was terminated after incubation at room temperature for 15 min. The labeled samples from each group were mixed equally. After removing the SDC, the supernatant was transferred to a new EP tube to obtain labeled polypeptide samples. After polypeptide desalting, rP-RP grading, and vacuum drying, frozen storage at -80 °C was used for LC-MS/MS analysis.

### 2.4.3 Proteomic data analysis

MaxQuant (version 1.6.1.0) was used for database search and TMT quantitative analysis of the original data obtained after LC-MS/MS analysis, while iBAQ non-standard quantitative analysis was performed. The UniProt_RATTus_20190711_ISO protein database was used. Polypeptide levels and protein false discovery rate (FDR) were controlled at 0.01. The 10 samples were then standardized such that the total protein or median was consistent in each group. DEPs were defined as proteins with a fold change (FC) >1.2 or <1/1.2, $P<0.05$, and a unique peptide≥2. Protein-protein interaction (PPI), Gene Ontology (GO), and Kyoto Encyclopedia of Genes and Genomes (KEGG) pathway enrichment analyses were performed on the DEPs.

### 2.5 Bioinformatics analysis of differentially expressed proteins (DEPs)

The PPI network was constructed using the STRING database (https://string-db.org/), and the analysis results in TSV format were downloaded and imported into the Cytoscape software to conduct further visual analysis of PPI [29]. Gene symbols corresponding to the DEPs were imported into the Kobass database (http://kobas.cbi.pku.edu.cn/), and the species *Rattus norvegicus* was selected for KEGG pathway enrichment analysis of DEPs [30]. The Cytoscape plug-in BiNGO was used to perform GO enrichment analysis from three aspects: biological process (BP), molecular function (MF), and cellular component (CC), to describe the properties of genes and gene products in an organism [31].

### 2.6 Western blotting (WB) protocol

A section of the bladder neck (50 mg) was added to 500 μL of RIPA cracking solution, mixed evenly, homogenized with a homogenizer (15 min), and placed on ice for cracking for 20 min, then centrifuged at 4 °C, 12000 rpm, for 20 min. The supernatant was transferred into an EP tube, protein concentration was measured using the BCA method, and a sample solution was prepared. It was placed in a metal bath for 15 min. SDS-PAGE glue (10%) was prepared for protein isolation by electrophoresis, and the proteins on the SDS-PAGE glue were electrotransferred to a polyvinylidene difluoride (PVDF) membrane. The PVDF membrane was soaked in the sealing solution containing 5% skim milk powder and incubated at room temperature for 1 h. TBST cleaning was performed three times for 10 min each. The primary antibodies six transmembrane epithelial antigen 3 (STEAP3) (1:1000; Abcam; ab151566), transferrin (TF) (1:10,000; Proteintech Group, Inc; 66171-1-lg), transferrin receptor (TFRC) (1:1000; Abcam; ab80194), divalent metal transporter1 (DMT1) (1:1000; Proteintech Group, Inc; 20507-1-AP), P53 (1:10000; Proteintech Group, Inc; 60283-2-lg), and smoothelin (SMTN) (1:1000; ABclonal; A6745) were added and incubated at 4 °C overnight; TBST cleaning was performed three times. The second antibody was added 10 min later: anti-mouse IgG (1:10000; Elabscience Biotechnology Co., Ltd.; E-AB-1001), Anti-Rabbit IgG (1:10,000; Elabscience Biotechnology Co., Ltd.; E-AB-1003), and incubated at room temperature for 1 h. TBST cleaning was performed three times for 10 min each and added developer on PVDF membrane for exposure and development in a dark room.

**2.7 Ferrous ion concentration detection**

Fresh bladder neck tissue (0.1 g) was taken, 0.9 ml buffer homogenate was added, the homogenate was centrifuged at 10000g for 10 min, and the supernatant was collected for use. Iron standards with different concentration gradients were added to the corresponding 1.5 mL EP tubes with 300 μL of different concentrations and the samples to be tested, and then 1.5 μL of the color-developing solution was added to each tube. After mixing well, the tubes were incubated at 37 °C for 10 min and centrifuged at 12000g for 10 min. The 300 μL supernatant was added to the enzyme label plate, and the optical density (OD) values of each well were measured at 593 nm using an enzyme label instrument. A ferrous ion colorimetric test kit was purchased from Elabscience Biotechnology Co., Ltd. (E-BC-K773-M).

**2.8 Masson's trichome staining**

Paraffin sections were dewaxed in water, soaked overnight in potassium dichromate, soaked in iron hematoxylin dye solution (Servicebio, China) for 3 min, differentiated in differentiation solution (Hunan Aifang Biotechnology Co., Ltd.), blued back in blue solution (Hunan Aifang Biotechnology Co., Ltd.), and soaked in Lichun Red acid fuchsin for 5–10 min. The samples were stained with phosphomolybdic acid for 1–3 min, aniline blue for 3–6 min, differentiated by 1% glacial acetic acid, dehydrated by anhydrous ethanol, sealed transparently, and the deposition of collagen fibers in the bladder neck was observed under a microscope.

**2.9 Statistical analysis**

Statistical software (SPSS 26.0) was used for analysis, and all data were expressed as mean ± standard deviation (x±s). One-way analysis of variance (ANOVA) was used for comparisons between groups, and a t-test was used for pairwise comparisons. $P<0.05$ was considered statistically significant.

## 3. Results

### 3.1 Electroacupuncture can improve the urination function of the neurogenic bladder after suprasacral spinal cord injury

The sham operation group was generally in good condition. In the Model and Model+EA groups, all voluntary hind limb movements ceased after spinal shock, and the limbs were dragged when walking. The bladder was enlarged, and the "olive" bladder could be touched in the lower abdomen. The lower abdomen of the rat and the cushion material in the cage were mildly wet, and resistance to the urethral orifice was experienced during manual urination. After electroacupuncture treatment, the degree of bladder distension was reduced, resistance to manually assisted urination was reduced, and urine was easily excreted.

The urodynamic results showed that compared with the sham operation group, LPP and MCC in the Model group were significantly increased ($P<0.01$). Compared with the Model group, LPP and MCC in the Model+EA group were significantly reduced after electroacupuncture intervention ($P<0.01$) (Figures 1A and 1B).

HE staining showed that the bladder neck tissue of the sham group was clearly arranged without inflammatory cell infiltration, and some areas had a slight hematoma. Edema was observed in the submucosa of the bladder neck in the Model group, a large number of inflammatory cells infiltrated, and collagen fiber hyperplasia was observed in the lamina propria. Compared with the Model group, submucosal edema in the bladder neck of the Model+EA group was reduced. There was no obvious inflammatory cell infiltration or collagen fiber hyperplasia in the lamina propria and smooth muscle fibers were neatly arranged (Figure 1C).

The results showed that electroacupuncture could improve the urination function of NB after SSCI.

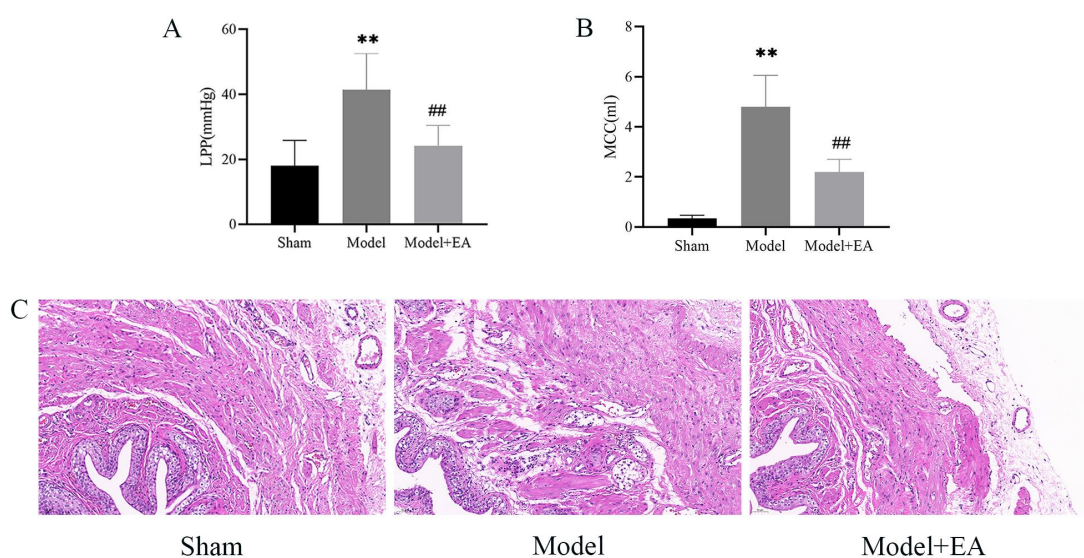

Figure 1 (A) and (B) show the comparison of leak point pressure and maximum cystometric capacity between the three groups, respectively. * Compared with the sham group, $P<0.05$; ** compared with the sham group, $P<0.01$; # Compared with the Model group, $P<0.05$; ## compared with the Model group, $P<0.01$. (C) HE staining images of the bladder neck (×200).

## 3.2 Quantitative proteomic analysis and verification of Tandem mass tag (TMT)
### 3.2.1 Tandem mass tag (TMT) analysis of differentially expressed proteins (DEPs)

In total, 4390 quantifiable proteins were detected using TMT quantitative proteomics. The DEPs of the Model/sham and Model+EA/Model groups were screened according to FC>1.2 or <1/1.2 ($P<0.05$) and unique peptide≥2, respectively. Among them, there were 614 DEPs in the Model/sham group (306 upregulated, 308 downregulated) and 145 DEPs (56 upregulated, 79 downregulated) in the Model+EA/Model group (Figures 2A and 2B). Comparative analysis of the DEPs between the Model/sham and Model+EA/Model groups revealed 59 overlapping DEPs (Figure 2C).

### 3.2.2 Protein-protein interaction (PPI) analysis was performed for the first 34 differentially expressed proteins (DEPs)

The PPIs of the 34 DEPs were mapped using the STRING online database and Cytoscape software. Each node represented a DEP. The darker the node color, the closer the DEPs are connected. Among them, it was found that STEAP3 and SMTN were associated many times; STEAP3 was the key protein in the ferroptosis signal pathway, and SMTN was the marker of contractile smooth muscle cells (Figure 2D).

### 3.2.3 Kyoto Encyclopedia of Genes and Genomes (KEGG) pathway analysis of differentially expressed proteins (DEPs)

KEGG pathway analysis of 34 DEPs using KOBAS revealed thyroid hormone synthesis, protein processing in the endoplasmic reticulum, ribosome, protein export, and prion. Seventeen signaling pathways, including diseases, ferroptosis, N-glycan biosynthesis, the p53 signaling pathway, and other signaling pathways, were significantly enriched, and the ferroptosis signaling pathway ranks the forefront (Figure 2E).

### 3.2.4 Gene ontology (GO) annotations of differentially expressed proteins (DEPs)

GO functions (BP, CC, and MF) were annotated using the ClueGO plug-in in Cytoscape. BP focuses on cellular response to stress and regulation of the apoptotic process. CC mainly focuses on intracellular organelle, endomembrane system, and endoplasmic reticulum. MF mainly focuses on binding, protein binding, catalytic activity, and receptor binding (Figure 2F).

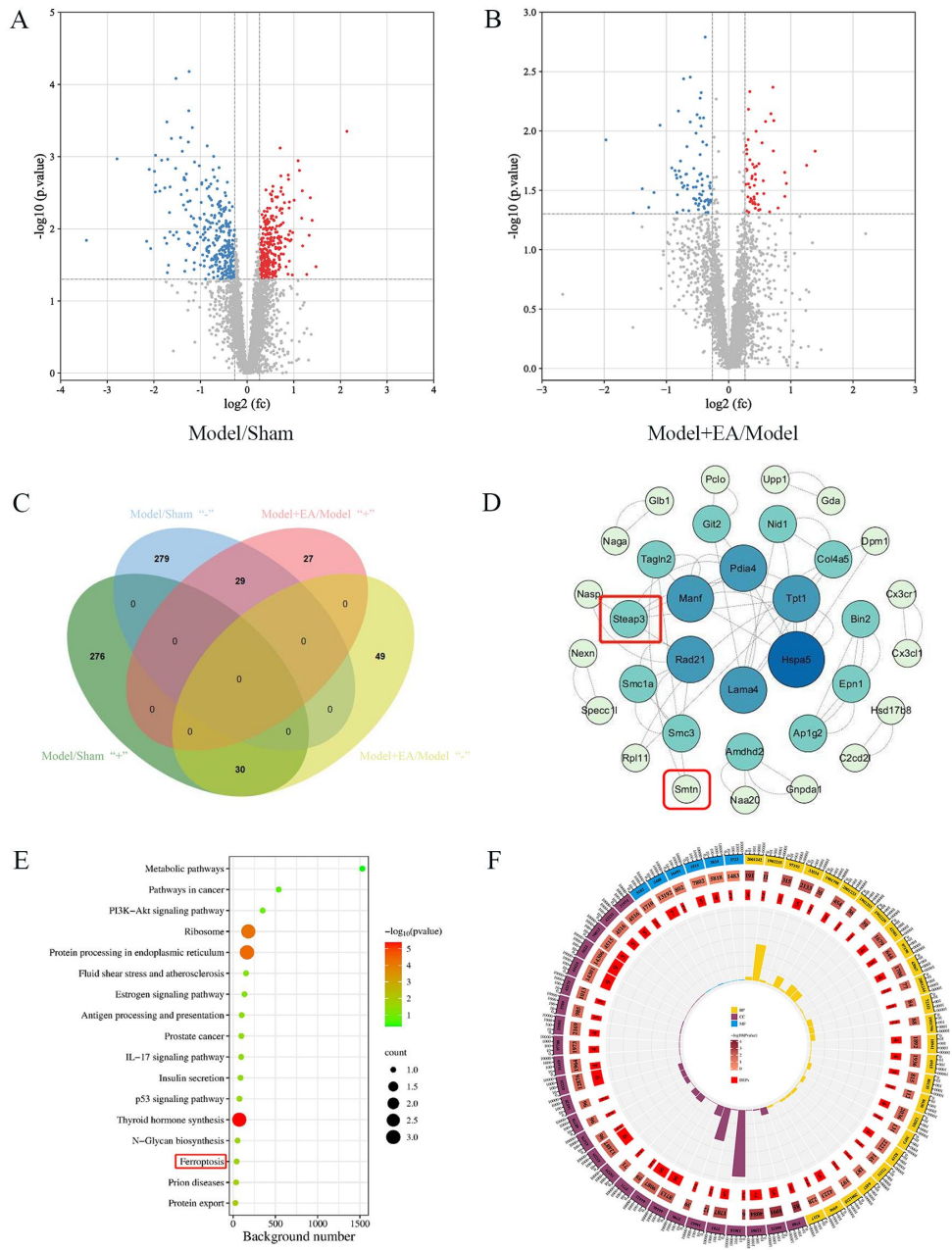

Figure 2 Quantitative proteomic analysis of TMT. (A) Model/sham's volcano map, which contains 614 DEPs. (B) Model+EA/Model A map of the volcano, which contains 145 DEPs. (C) Venn diagrams of DEPs and 59 overlapping proteins, with "+" representing upregulated DEPs and "-" representing downregulated DEPs. (D) PPI maps for 34 DEPs. (E) KEGG pathway maps for 34 DEPs. The horizontal coordinate is the background value of pathway enrichment; the size of the point is the number of genes; the vertical coordinate is the P-value of pathway enrichment; and the left side is the name of the gene involved in the pathway. (F) 34 DEPs GO feature annotations. Where yellow represents BP, purple represents CC, and blue represents MF.

**3.2.5 Western blotting verification of STEAP3, a key protein related to ferroptosis, which was enriched in protein-protein interaction (PPI) analysis**

To further verify the reliability of the TMT quantitative proteomic results, we performed WB validation on the STEAP3 protein enriched in the PPI analysis. The results showed that compared with the sham group, STEAP3 expression in the bladder neck of the Model group was increased ($P<0.05$). Compared with the Model group, the expression of STEAP3 in the bladder neck of the Model+EA group decreased after electroacupuncture intervention ($P<0.05$), and the results were consistent with the quantitative proteomic data of TMT (Figure 3).

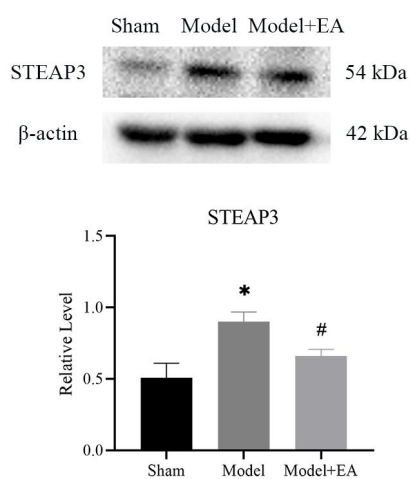

Figure 3 Quantitative proteomic results of TMT were verified by WB. The results of WB were consistent with those of TMT quantitative proteomics. *$P<0.05$, compared with sham group; # $P<0.05$, compared with the Model group.

### 3.3 Electroacupuncture effects on ferroptosis of the bladder neck
### 3.3.1 Western blotting verification of ferroptosis signaling pathway-related proteins

To further confirm the important role of ferroptosis in the pathological changes of the bladder neck, we selected four key proteins from the ferroptosis signaling pathway for WB verification. The results showed that compared with the sham group, the expression of TF, TFRC, DMT1, and P53 in the bladder neck of the Model group was increased ($P<0.05$). Compared with the Model group, the expression of TF, TFRC, DMT1, and P53 in the bladder neck of the Model+EA group decreased after electroacupuncture intervention ($P<0.05$) (Figure 4A–4D).

### 3.3.2 Ferrous ion concentration detection

Ferroptosis alters the concentration of ferrous ions in the body. Compared with the sham group, the concentration of ferrous ions in the model group was significantly higher ($P<0.01$). Compared with the Model group, the concentration of ferrous ions in the Model+EA group was significantly lower ($P<0.01$) (Figure 4E).

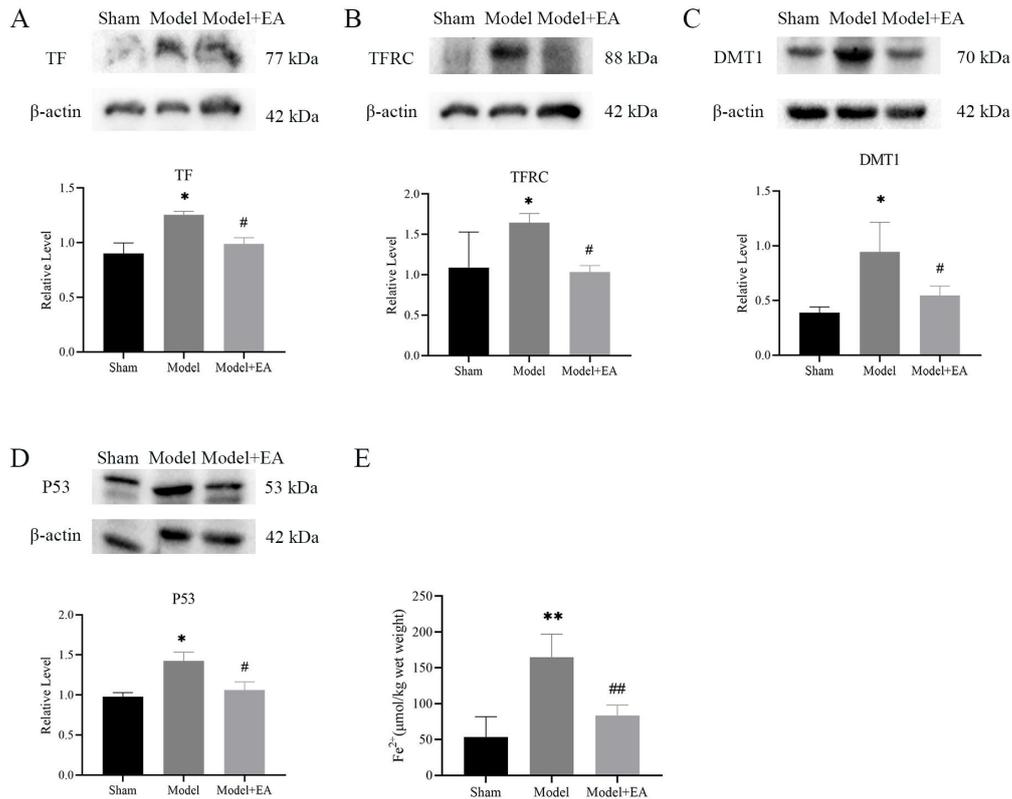

Figure 4 (A) – (D) The expression of related proteins in the ferroptosis signaling pathway was verified by WB; (E) Comparison of ferrous ion concentration in bladder neck among different groups. *$P<0.05$, compared with the sham group; **$P<0.01$, compared with the sham group; # $P<0.05$, compared with the Model group; ## $P<0.01$, compared with the Model group.

### 3.4 Effects of electroacupuncture on bladder neck fibrosis
### 3.4.1 Results of Masson's trichome staining for bladder neck fibrosis

Compared with the sham group, the blue-dyed fibers of the bladder neck muscle layer in the Model group were significantly increased, suggesting a large amount of collagen fiber deposition. Compared with the Model group, the blue-dyed fibers of the bladder neck muscle layer were significantly reduced in the Model+EA group. According to the statistical analysis using ImageJ software, the degree of bladder neck fibrosis in the Model group was significantly higher than that in the sham group ($P<0.01$), whereas the degree of bladder neck fibrosis in the Model+EA group was lower than that in the Model group ($P<0.01$) (Figures 5A and 5B).

### 3.4.2 Western blotting verification of contractility smooth muscle cell marker SMTN

The results showed that, compared with the sham group, the expression of SMTN in the Model group was decreased ($P<0.05$). Compared with the Model group, the expression of SMTN in the Model+EA group increased after electroacupuncture intervention ($P<0.05$) (Figure 5C).

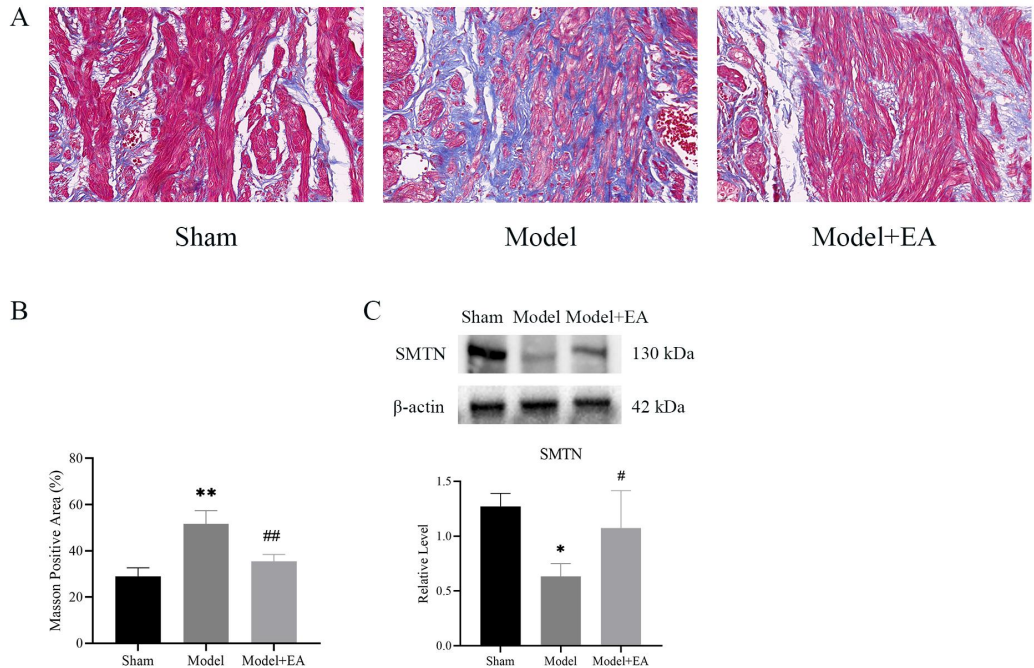

Figure 5 (A) Masson's trichome staining of the bladder neck (×400), with blue representing collagen fibers and red representing muscle fibers. (B) Image J software was used to analyze the results of Masson's trichome staining of bladder neck tissue and calculate the percentage of collagen fiber-positive area. (C) WB results of SMTN. *$P<0.05$, compared with the sham group; **$P<0.01$, compared with the sham group; # $P<0.05$, compared with the Model group; ## $P<0.01$, compared with the Model group.

## 4. Discussion

NB is a major clinical challenge after spinal cord injury, and its clinical mechanism mainly lies in overexcitation of the sympathetic nerve, continuous contraction of the bladder neck, and increased internal bladder pressure, resulting in tissue ischemia and hypoxia, which promote the phenotypic transformation of bladder neck smooth muscle cells, transforming them from contractile to synthetic, secreting a large amount of extracellular matrix, and causing fibrosis [25]. Consequently, bladder neck compliance is reduced, resulting in excessive pressure at the urethral orifice and a sharp increase in bladder pressure during urination, leading to urinary reflux and renal function injury. The role of the bladder neck in the urination process should not be ignored, and its use as a target for NB therapy is of great significance.

There are many records of acupuncture treatment for this disease in ancient Chinese medicine, and there is no lack of research on its use in modern medicine. "Ciliao (BL32)" is located in the second sacral posterior foramina, under which sacral nerves pass and play a role in regulating the bladder neck, causing the bladder to contract rhythmically and relax the bladder muscle [32, 33]. Zhongji (RN3)" is located near the projection area of the bladder body surface and is innervated by the T12-L1 segment of the spinal cord, which overlaps with the spinal segment innervating the sympathetic nerve of the bladder; therefore, acupuncture at this point can affect the bladder neck [28, 34]. "Sanyinjiao (SP6)" is controlled by the S2 segment of the spinal cord and is related to the sacral pulp urination center, which regulates bladder compliance [35]. Previous studies by our research group found [34] that electroacupuncture at these three points positively affects NB. Therefore, we used electroacupuncture as an intervention method to prepare an NB rat model after T10 spinal cord transsection injury and placed the target on the bladder neck to carry out this experiment.

In the urodynamic examination, we chose MCC and LPP as the main effect indices. The bladder was perfused with normal saline to simulate urination. With an increase in perfusion, the bladder volume and pressure increase. Due to dyssynergia of the bladder neck, the sphincter of the bladder neck is constantly tense, and the bladder neck cannot be relaxed and opened. Effective urination can only be achieved when the perfusion volume increases to make the internal bladder pressure higher than the bladder neck pressure; the intravesical pressure during urination is the LPP. Therefore, LPP reflects the pressure of the bladder neck and the tension of the bladder neck sphincter. Simultaneously, the greater the LPP, the greater the bladder capacity, and both were positively correlated. The results showed that compared with the sham group, the LPP and the MCC of the Model group were significantly increased, which indicated that the bladder neck sphincter was excessively contracted after the shock stage of SSCI, leading to a sharp increase in internal bladder pressure and an increase in bladder capacity. Compared with the Model group, the LPP and MCC of rats in the Model+EA group were significantly reduced. This shows that electroacupuncture can effectively relieve excessive tension of the bladder neck, reduce urethral orifice pressure during urination, and solve the problem of increased bladder capacity caused by urine not being discharged in time, thus improving bladder urination function.

HE staining showed that, compared with the sham group, the Model group had edema, many inflammatory cells infiltrated, and the arrangement of muscle fibers was disordered. Compared with the Model group, the histopathological lesions of the bladder neck in the Model+EA group improved to a certain extent; edema of the submucosa was reduced; there was no obvious inflammatory cell infiltration or collagen fiber hyperplasia in the lamina propria; and smooth muscle fibers tended to be arranged neatly. These results are consistent with those of previous studies [34, 36, 37], which confirmed that electroacupuncture could effectively improve inflammation and other pathological changes in the bladder neck tissue after SSCI.

To further explore the potential targets and mechanisms of action in bladder neck tissues of rats with NB after electroacupuncture treatment of SSCI, we extracted bladder neck tissues and used TMT quantitative labeling proteomics to screen out DEPs in the bladder neck. We found that there were 59 overlapping differentially expressed proteins between the Model/Sham and Model+EA/Model, and all of them were inversely regulated by electroacupuncture. We then performed bioinformatics analysis on these differentially expressed proteins. Using the STRING online database combined with Cytoscape software for protein interaction analysis, we found that 34 differentially expressed proteins were closely connected and formed a large cluster. KEGG enrichment analysis was performed for these 34 differentially expressed proteins, and 17 pathways were significantly enriched. We focused on the pathways and the related differentially expressed proteins on the pathways that may affect bladder neck compliance, and found that the ferroptosis, p53 signaling pathway, and STEAP3 proteins on the ferroptosis signaling pathway are significantly enriched in them. STEAP3 is an iron reductase [38, 39] belonging to the STEAP family, which is homologous to oxidoreductase found in bacteria and archaea and FRE metal reductase in yeast [40]. It exhibits iron reductase activity, reducing trivalent iron in endosomes to ferrous iron [40, 41]. As STEAP3 plays a key regulatory role in ferroptosis by mediating iron metabolism [42, 43], we conducted WB verification of STEAP3. The results showed that the expression of STEAP3 in the bladder neck decreased after electroacupuncture treatment, which was consistent with the results of TMT proteomics.

Ferroptosis is a cell death mechanism mainly involved in iron metabolism and lipid peroxidation [43, 44]. Its morphological characteristics differ from those of apoptotic cell death, which includes cell contraction, nuclear fragmentation, and chromatin aggregation. Ferroptotic cells exhibit necrotic morphology, such as incomplete plasma membrane rupture and release of cell contents [45], which are inextricably related to iron metabolism and iron homeostasis. Although physiological iron levels can promote cell growth, instability in iron redox may affect cellular sensitivity to ferroptosis [46]. When the $Fe^{2+}$ concentration increases in the cell, hydrogen peroxide combines with it to produce a Fenton reaction, forming hydroxides and hydroxyl radicals, thereby increasing oxidative damage [47, 48]. Hydroxyl radicals are important reactive oxygen species (ROS). Because iron plays an important role in ROS production, ferroptosis is regulated by multiple aspects of iron metabolism, including iron uptake, storage, utilization, and efflux. The present study suggests that the P53 protein and its p53 signaling pathway also play a role in regulating ferroptosis. To further confirm the important role of ferroptosis in the pathological changes of the bladder neck, we detected the proteins TF, TFRC, DMT1, and P53 in related pathways, as well as the intracellular $Fe^{2+}$ concentration. The results showed that the expression of TF, TFRC, DMT1, and P53 in the bladder neck decreased after electroacupuncture treatment. Electroacupuncture also significantly reduced the concentration of free ferrous ions in the bladder neck, confirming that electroacupuncture may effectively inhibit the ferroptosis signaling pathway in the bladder neck of NB model rats.

Under normal circumstances, the body obtains iron through diet, and the non-heme iron in food is mainly insoluble $Fe^{3+}$, which must be reduced to $Fe^{2+}$ for absorption. $Fe^{3+}$ binds to TF in serum and is recognized by TFRC on the cell membrane to form a TFRC/TF-$(Fe^{3+})_2$ complex, which enters the cell as an endosome. Lysosomal $Fe^{3+}$ is reduced to $Fe^{2+}$ by prostatic STEAP3 and finally released from the endosome into the cytoplasm via solute vector family 11 member 2 (SLC11A2, best known as DMT1). In the ferroptosis signaling system, TF is the carrier protein of iron in the serum, and two molecules of $Fe^{3+}$ can be combined with one molecule of TF, transported to cells through TFRC-mediated endocytosis in an alkaline environment, and acidified by the acidic environment of the intracellular fluid so that the complex dissociates $Fe^{3+}$[49]. DMT1 is a transmembrane glycoprotein and a major iron transporter that can transport ferrous but not trivalent iron. Excess ferrous iron, when excreted, may damage cells and tissues through the production of ROS.

Due to the regulatory function of P53 in many basic cells, its protein levels and activity are strictly regulated. Under non-stress conditions, P53 is maintained at low levels in normal cells; however, due to its significantly increased half-life, P53 accumulates in response to various intracellular and extracellular stress signals, such as hypoxia, oncogene activation, and ferroptosis. Once activated, P53 binds to P53-responsive elements in its target genes to transcriptionally regulate expression [50–53]. It has been reported that P53 sensitizes cells to ferroptosis via the transcriptional inhibition of SLC7A11 expression. SLC7A11 is a direct target of P53, and its expression is inhibited by binding to p11 response elements in the SLC53A7 promoter region [54]. The cystine glutamate transporter (Xc-system) is a heterodimer composed of two subunits, SLC7A11 and SLC3A2, which are the main carriers of cystine and glutamate exchange in and out of cells. Intracellular cysteine is mainly derived from the breakdown of cystine; therefore, this receptor greatly affects intracellular cysteine levels. Cysteine is an important raw material for synthesizing glutathione (GSH), and its reduction can significantly reduce intracellular GSH content. GSH can produce a reducing effect under the action of glutathione peroxidase 4 (GPX4) and negatively regulate lipid peroxide content in the cell. Therefore, P53 can indirectly reduce the GSH substrate of GPX4 and induce ferroptosis in bladder neck smooth muscle cells [55]. Some studies have found that P53 can directly regulate STEAP3 and that there is a P53 binding site in the STEAP3 promoter, which can upregulate the expression of STEAP3 and promote cell death when P53 is activated [56, 57]. Therefore, the P53 protein also plays a crucial role in the process of ferroptosis, and reducing the level of P53 in cells can effectively inhibit the occurrence of iron death, as shown in Figure 6.

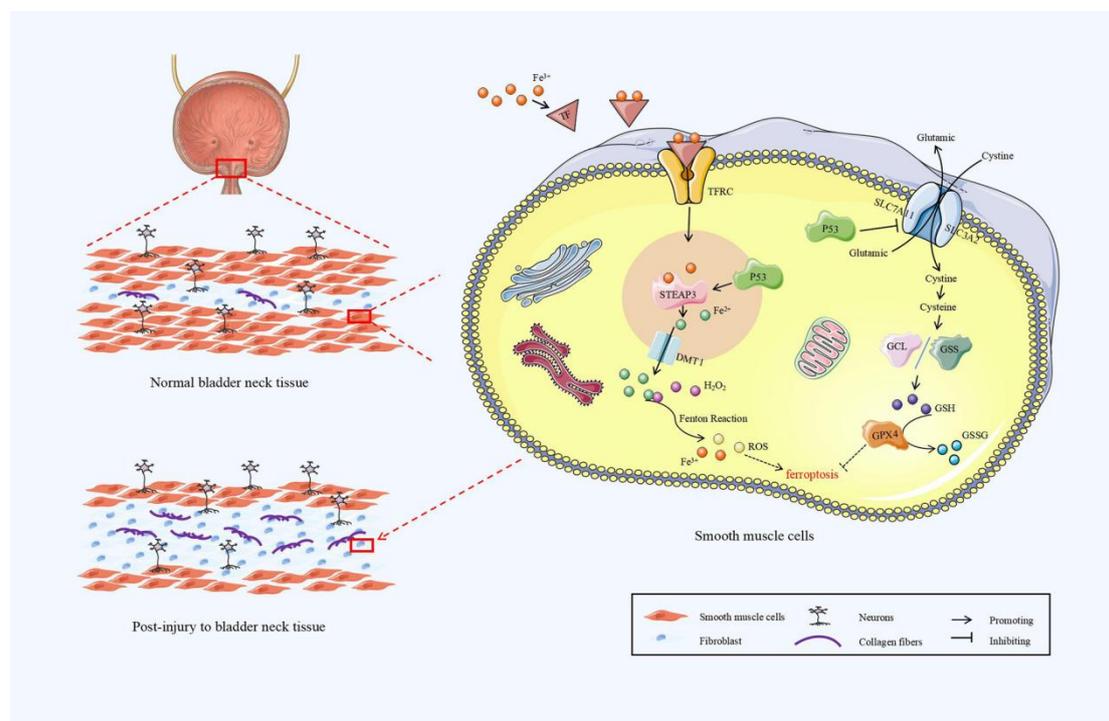

Figure 6 Ferroptosis associated with fibrosis in bladder neck smooth muscle cells

Studies have found that ferroptosis plays an important role in the phenotypic transformation of smooth muscle cells and can promote the transformation of smooth muscle cells from the contractile to the synthetic type. Excess ROS produced during ferroptosis can damage the redox balance of smooth muscle cells, which is a key factor in inducing phenotypic transformation [20, 21]. During the transformation of smooth muscle cells from the contractile type to the synthetic type, a large amount of extracellular matrix is secreted, which is the main reason for the fibrosis of the smooth muscle of the bladder neck [25] and is an important factor leading to the reduction of compliance in the bladder neck. Masson's trichome staining was performed on bladder neck tissue to confirm this effect, and WB was performed to verify SMTN, a specific marker of contractile smooth muscle cells. The results of Masson's trichome staining showed that, compared with the sham group, the number of blue-dyed fibers in the Model group increased significantly, suggesting that the deposition of collagen fibers increased. Compared with the Model group, the Model+EA group showed significantly reduced deposition of collagen fibers in the bladder neck muscle layer and improved fibrosis of the bladder neck. WB results showed that the expression of SMTN in the bladder neck increased after electroacupuncture treatment, confirming that electroacupuncture can inhibit the phenotypic transformation of smooth muscle cells and prevent fibrosis by regulating ferroptosis.

This study analyzed the expression of DEPs and enrichment of KEGG signal transduction pathways using TMT proteomics technology. It was found that electroacupuncture may protect the smooth muscle cells of bladder neck contraction phenotype, reduces tissue fibrosis, improves bladder neck compliance and restores bladder relaxation through ferroptosis and the p53 signaling pathway. Electroacupuncture can effectively inhibit the occurrence of ferroptosis in the bladder neck and prevent smooth muscle cells from transforming into synthetic cells, thereby protecting the compliance of the bladder neck and enabling it to relax spontaneously during urination.

In this study, WB, a biochemical kit, and Masson's trichome staining were used to verify the mechanism related to ferroptosis. However, agonists or inhibitors were not set as key targets in the ferroptosis signaling pathway. We plan to improve the scheme and conduct further studies.

**Acknowledgments**

We thank the Animal Laboratory Center of Hunan University of Traditional Chinese Medicine and the Key Laboratory of Hunan Province for Integrated Traditional Chinese and Western Medicine on Prevention and Treatment of Cardio-Cerebral Diseases for their support and assistance. We would like to thank Editage (www.editage.cn) for English language editing.

**Funding**

We acknowledge the Natural Science Foundation of China (General) (No.81874510), National Natural Science Foundation Youth Project (No.82205255), Natural Science Foundation of Hunan Province (No.2022JJ40301), Hunan University of Chinese Medicine graduate student innovation project (No.2023CX74).

**Availability of data and materials**

The data generated in the present study may be requested from the corresponding author

**Author Contributions**

Study design: MX, KA. project administration: MX, LQ. Experiment implementation: JCL, LYT, QRQ. Data analysis: JCL, YXS, LZ. Paper writing: JCL, LYT, KA. Paper review & editing: HZ. Experimental support: MX, KA. All authors approved the final version of the manuscript.

**Ethics Approval and Consent to Participate**

Experimental animal ethics committee of Hunan University of Chinese Medcine. Ethical approval number: LL2019092303.

**Competing interests**

The authors declare no competing financial interest.